# Functional observers with linear error dynamics for discrete-time nonlinear systems*

Sunjeev Venkateswaran, Benjamin A. Wilhite, Costas Kravaris

**Abstract— This work deals with the problem of designing observers for the estimation of a single function of the states for discrete–time nonlinear systems. Necessary and sufficient conditions for the existence of lower order functional observers with linear dynamics and linear output map are derived. The results provide a direct generalization to Luenberger's linear theory of functional observers. The design methodology is tested on a non-isothermal CSTR case study.**

## I. Introduction

In control theory, a functional observer is an auxiliary system that is driven by the available system outputs and mirrors the dynamics of a physical process in order to estimate one or more functions of the system states [1, 2]. Besides being of theoretical importance, the use of functional observers arises in many applications. For example, functional estimates are useful in feedback control system design because the control signal is often a linear combination of the states, and it is possible to utilize a functional observer to directly estimate the feedback control signal[1-3].

Over the past fifty years, considerable research has been carried out on estimating functions of the state vector for linear systems ever since Luenberger introduced the concept of functional observers in 1966[2] and proved that it is feasible to construct a functional observer with number of states equal to observability index minus one. Subsequent research has focused on lower order functional observers where necessary and sufficient conditions for their existence and stability have been derived[4-6], and parametric approaches to the design of lower order functional observers[7] and algorithms for solving the functional observer design conditions have also been developed[4, 5, 8]

For nonlinear systems, functional observers for Lipschitz systems [9, 10] and a class of nonlinear systems that can be decomposed as sum of Lipschitz and non-Lipschitz parts [7] (with the non-Lipschitz part considered as an unknown input/disturbance) have been developed. More recently, the problem of designing functional observers for estimating a single nonlinear functional has been tackled for general nonlinear systems from the point of view of observer error linearization[3] and the approach has been extended to a disturbance decoupled fault detection and isolation[11]. For discrete-time nonlinear systems, however, results have been limited. The goal of the present work is to develop a direct generalization of Luenberger's functional observers to discrete time nonlinear systems. The concept of functional observers for discrete-time nonlinear systems is defined and the observer design problem is considered from the point of view of observer error linearization and is analogous to the methods in [3, 11, 12]. It will be shown that, with the proposed formulation, easy-to-check necessary and sufficient conditions for the existence of such a functional observer can be derived, leading to simple formulas for observer design with eigenvalue assignment. Furthermore, the formulation also lends itself to fault detection and estimation in discrete-time nonlinear systems and this will also be investigated.

The outline of this present study is as follows. In the next couple of sections, the notion of functional observer for discrete time nonlinear systems will be defined in a manner completely analogous to Luenberger's definition[1, 2] for linear systems and different approaches to solve the functional observer design problem will be outlined. Following this, notions of observer error linearization will be defined, and then necessary and sufficient conditions will be derived for the solution of the linearization problem, as well as a simple formula for the resulting functional observer.

## II. FUNCTIONAL OBSERVERS FOR DISCRETE-TIME NONLINEAR SYSTEMS

Consider a discrete- time nonlinear system described by:

$$
\begin{aligned}
x(k+1) &= F\big(x(k)\big) \qquad \text{(II.1)}\\
y(k) &= H\big(x(k)\big)\\
z(k) &= q(x(k))
\end{aligned}
$$

where:

$x \in \mathbb{R}^n$ is the system state

$y \in \mathbb{R}^p$ is the vector of measured outputs

$z \in \mathbb{R}$ is the (scalar) output to be estimated

and $F: \mathbb{R}^n \to \mathbb{R}^n, H: \mathbb{R}^n \to \mathbb{R}^p, q: \mathbb{R}^n \to \mathbb{R}$ are smooth nonlinear functions. The objective is to construct a functional observer of order $\nu < n$, which generates an estimate of the output z, driven by the output measurement y.

In complete analogy to Luenberger's construction for the linear case, we seek a mapping

*Research supported National Science Foundation through the grant CBET-1706201.All authors are with Texas A&M University, College Station, TX, 77843, USA (corresponding author e-mail: kravaris@tamu.edu).

.

$$\xi = T(x) = \begin{bmatrix} T_1(x) \\ \vdots \\ T_\nu(x) \end{bmatrix}$$

from $\mathbb{R}^n$ to $\mathbb{R}^\nu$, to immerse system (2.1) to a $\nu$ -th order system ($\nu < n$), with input y and output z:

$$\xi(k+1) = \varphi(\xi(k), y(k)) \quad (2.2)$$
$$z(k+1) = \omega(\xi(k), y(k))$$

But in order for system (2.1) to be mapped to system (2.2) under the mapping T(x), the following relations have to hold

$$\varphi(T(x), H(x)) = T\big(F(x)\big) \quad (2.3)$$

$$\omega\big(T(x), H(x)\big) = q(x) \quad (2.4)$$

The foregoing considerations lead to the following definition of a functional observer:

Definition 1: Given a dynamic system

$$x(k+1) = F\big(x(k)\big) \quad (2.1)$$
$$y(k) = H(x(k))$$
$$z(k) = q(x(k))$$

where $F: \mathbb{R}^n \to \mathbb{R}^n, H: \mathbb{R}^n \to \mathbb{R}^p, q: \mathbb{R}^n \to \mathbb{R}$ are smooth nonlinear functions, y is the vector of measured outputs and z is the scalar output to be estimated, the system

$$\hat{\xi}(k+1) = \varphi(\hat{\xi}(k), y(k)) \quad (2.5)$$
$$\hat{z}(k+1) = \omega(\hat{\xi}(k), y(k))$$

if in the series connection

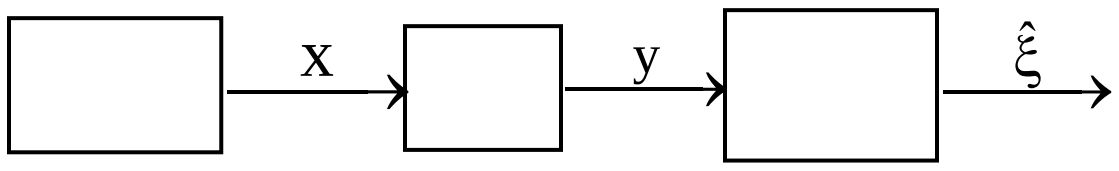

the overall dynamics

$$x(k+1) = F(x(k))$$
$$\hat{\xi}(k+1) = \varphi(\hat{\xi}(k), H(x(k)))$$

possesses an invariant manifold $\hat{\xi} = T(x)$ with the property that $q(x) = \omega(T(x), H(x))$.

If the functional observer (2.5) is initialized consistently with the system (2.1) i.e. if

$\hat{\xi}(0) = T(x(0))$, then $\hat{\xi}(k) = T(x(k)),\ \forall\, k \in \mathbb{N}$ and therefore

$\hat{z}(k) = \omega\left(\hat{\xi}(k), y(k)\right) = \omega\left(T\big(x(k)\big), H\big(x(k)\big)\right) = q\big(x(k)\big) \quad \forall\, k \in \mathbb{N}$ , the functional observer will be able to exactly reproduce z(k).

In the presence of initialization errors, additional stability requirements will need to be imposed on the $\hat{\xi}$-dynamics, for the estimate $\hat{z}(k)$ to asymptotically converge to z(k).

At this point, it is important to examine the special case of a linear system, where $F(x) = Fx, H(x) = Hx, q(x) = qx$ with F, H, q being $n \times n, p \times n, 1 \times n$ matrices respectively, and a linear mapping $T(x) = Tx$ is considered. Definition 1 tells us that for a linear time-invariant system

$$x(k+1) = Fx(k) \quad (2.6)$$
$$y(k) = Hx(k)$$

$$z(k) = qx(k)$$

the system

$$\hat{\xi}(k+1) = A\hat{\xi}(k) + By(k) \quad (2.7)$$
$$\hat{z}(k) = C\hat{\xi}(k) + Dy(k)$$

will be a functional observer if the following conditions are met:

$$TF = AT + BH$$

$$q = CT + DH$$

for some $\nu \times n$ matrix T. These are exactly the discrete-time version of Luenberger's conditions for a functional observer for linear continuous time-invariant systems [1, 2]

## III. DESIGNING LOWER ORDER FUNCTIONAL OBSERVERS

For the design of a functional observer, one must be able to find a continuous map $T(x) = \begin{bmatrix} T_1(x) \\ \vdots \\ T_\nu(x) \end{bmatrix}$ to satisfy conditions (2.3) and (2.4) i.e. such that $T_j\big(F(x)\big), j = 1, \cdots, \nu$ is a function of $T_1(x), \cdots, T_\nu(x), H(x)$ and $q(x)$ is a function of $T_1(x), \cdots, T_\nu(x), H(x)$

However, such scalar functions $T_1(x), \cdots, T_\nu(x)$ may not exist, if $\nu$ is too small. Moreover, even when they do exist, there is an additional very important requirement: Since

$T(F(x)) = \varphi(T(x), H(x))$ will define the right-hand side of the functional observer's dynamics, it must be such that the functional observer's dynamics is stable and the decay of the error is sufficiently rapid.

All the above requirements can be satisfied if $\begin{cases} x(k+1) = F(x(k)) \\ y(k) = H(x(k)) \end{cases}$ is linearly observable and $\nu = n - p$:

Available design methods for reduced-order state observers[13] generate a functional observer of order $\nu = n - p$, by only modifying the output map of the observer (so that the estimate of z is the observer output instead of the entire state vector).

The question is whether a lower order $\nu < n - p$ would be feasible and how to go about designing the functional observer. This is not an easy question because we will be trying to impose too many requirements at the same time.

For constructing the functional observer, one possible way involves identifying functions $T_1(x), \cdots, T_\nu(x)$ such that $T_j\big(F(x)\big), j = 1, \cdots, \nu$ and $q(x)$ can be expressed as functions of $T_1(x), \cdots, T_\nu(x)$ and the measured output. The second step is then to check stability of the error dynamics. This approach might be successful if the selection of $T_1(x), \cdots, T_\nu(x)$ could be directed by physical intuition.

Alternatively, one could try to follow the opposite path. As a first step, try to enforce stability: given some desirable dynamics for the observer $\hat{\xi}(k+1) = \varphi(\hat{\xi}, y)$, with $\varphi$ so as to guarantee stability and rapid decay of the error, find $T(x) = \begin{bmatrix} T_1(x) \\ \vdots \\ T_\nu(x) \end{bmatrix}$ so that $T(F(x)) = \varphi(T(x), H(x))$. The second step will then be to check if q(x) can be expressed as a function of $T(x)$ and $H(x)$.

## IV. EXACT LINEARIZATION OF A FUNCTIONAL OBSERVER

Along the second line of attack of the functional observer design problem, the most natural $\varphi$ - function to work with is the linear one:

$$\varphi(\xi, y) = A\xi + By$$

It will then be the eigenvalues of the matrix A that will determine stability of the functional observer and the rate of decay of the error.

If we can find a continuously differentiable map $T(x)$ to satisfy the corresponding condition (2.3), i.e. to be a solution of the functional equation

$$T(F(x)) = AT(x) + BH(x)$$

for some Hurwitz matrix A, and if in addition $T(x)$ satisfies condition (2.4), i.e. that q(x) can be expressed as a function of $T(x)$ and $H(x)$, then we have a stable functional observer with linear dynamics. This leads to the Functional Observer Linearization Problem.

Functional Observer Linearization Problem

Given a system of the form (2.1), find a functional observer of the form

$$\begin{aligned} \hat{\xi}(k+1) &= A\hat{\xi}(k) + By(k) \\ \hat{z}(k) &= C\hat{\xi}(k) + Dy(k) \end{aligned} \tag{4.1}$$

where A, B, C, D are $\nu \times \nu, \nu \times p, 1 \times \nu, 1 \times p$ matrices respectively, with A having stable eigenvalues. Equivalently, find a continuously differentiable mapping $T: \mathbb{R}^n \to \mathbb{R}^\nu$ such that

$$T(F(x)) = AT(x) + BH(x) \tag{4.2}$$

and

$$q(x) = CT(x) + DH(x) \tag{4.3}$$

Assuming that the above problem can be solved, the resulting error dynamics will be linear:

$$\hat{\xi}(k+1) - T(x(k+1)) = A\left(\hat{\xi}(k) - T\big(x(k)\big)\right)$$

$$\hat{z}(k) - z(k) = C\left(\hat{\xi}(k) - T\big(x(k)\big)\right) \tag{4.4}$$

from which $\hat{z}(k) - z(k) = CA^k\big(\hat{\xi}(0) - T(x(0))\big)$.

With the matrix A having eigenvalues in the interior of the unit disc, the effect of the initialization error $\hat{\xi}(0) - T(x(0))$ will die out, and $\hat{z}(k)$ will approach $z(k)$ asymptotically.

Remark 4.1: It is possible to formulate a linearization problem in a slightly more general manner by including additive nonlinear output injection terms in the functional observer and a possibly nonlinear output map

$$\begin{aligned} \hat{\xi}(k+1) &= A\hat{\xi}(k) + \mathcal{B}\big(y(k)\big) \\ \hat{z}(k) &= \omega(\hat{\xi}(k), y(k)) \end{aligned} \tag{4.5}$$

where $\mathcal{B}$ (.) $\mathbb{R}^p \to \mathbb{R}^\nu$is the nonlinear output injection term.

In order to solve the Functional Observer Linearization Problem, it is natural to first try to solve the system of functional equations (4.2) (see [14] and[15] for solvability conditions) given some small-size matrix A with fast enough eigenvalues, and then check to see if q(x) can be expressed as a function of $T(x)$ and $H(x)$ according to (4.3). If it can, we have a functional observer with linear error dynamics; if not, we can try a different matrix A with different eigenvalues and/or larger size, up until we can satisfy both conditions.

## V. NECESSARY AND SUFFICIENT CONDITIONS FOR SOLVABILITY OF THE FUNCTIONAL OBSERVER LINEARIZATION PROBLEM

The trial-and-error approach outlined in the previous section is in principle feasible, but it is far from being practical due to the computational effort involved in solving (4.2), which will be multiplied by the number of trials until (4.3) is satisfied.

To be able to develop a practical approach for designing functional observers, it would be helpful to develop criteria to check if for a given set of $\nu$ eigenvalues, there exists a functional observer whose error dynamics is governed by these eigenvalues. This will be done in the present Section for the Functional Observer Linearization Problem.

The main result is as follows:

*Proposition 1: Under the assumptions of Proposition 1, for a real analytic nonlinear system of the form (2.1), there exists a functional observer of the form*

$$\begin{aligned} \hat{\xi}(k+1) &= A\hat{\xi}(k) + By(k) \\ \hat{z}(k) &= C\hat{\xi}(k) + Dy(k) \end{aligned} \tag{4.1}$$

*with the eigenvalues of A being the roots of a given polynomial* $\lambda^\nu + \alpha_1\lambda^{\nu-1} + \cdots + \alpha_{\nu-1}\lambda + \alpha_\nu$,

*if and only if* $qF^\nu(x) + \alpha_1 qF^{\nu-1}(x) + \cdots + \alpha_{\nu-1}qF(x) + \alpha_\nu q(x)$ *is* $\mathbb{R}$*-linear combination of* $H_j(x), H_jF(x), \ldots, H_jF^\nu(x), j = 1, \cdots, p,$ *where in the above we have used the notation*

$F^j(x) = \underbrace{F \circ F \dots F \circ F(x)}_{j\ \text{times}}$ and $H_jF(x) = (H_j \circ F)(x)$

**Proof: Necessity:** Suppose that there exists $T(x) = \begin{bmatrix} T_1(x) \\ T_2(x) \\ \vdots \\ T_\nu(x) \end{bmatrix}$ such that (4.2) is satisfied, i.e

$$\begin{bmatrix} T_1F(x) \\ T_2F(x) \\ \vdots \\ T_\nu F(x) \end{bmatrix} = A \begin{bmatrix} T_1(x) \\ T_2(x) \\ \vdots \\ T_\nu(x) \end{bmatrix} + \begin{bmatrix} B_1H(x) \\ B_2H(x) \\ \vdots \\ B_\nu H(x) \end{bmatrix}$$

where $B_1, \dots, B_\nu$ denote the rows of the matrix B. Now, we find that for k=1,2,3…

$$\begin{bmatrix} T_1F^k(x) \\ T_2F^k(x) \\ \vdots \\ T_\nu F^k(x) \end{bmatrix} = A^k \begin{bmatrix} T_1(x) \\ T_2(x) \\ \vdots \\ T_\nu(x) \end{bmatrix} + \begin{bmatrix} (A^{k-1}B)_1H(x) + (A^{k-2}B)_1HF(x) + \cdots + (B_1HF^{k-1}(x)) \\ (A^{k-1}B)_2H(x) + (A^{k-2}B)_2HF(x) + \cdots + (B_2HF^{k-1}(x)) \\ \vdots \\ (A^{k-1}B)_\nu H(x) + (A^{k-2}B)_\nu HF(x) + \cdots + (B_\nu HF^{k-1}(x)) \end{bmatrix}$$

and we can calculate

$$T_iF^\nu(x) + \alpha_1T_iF^{\nu-1}(x) + \cdots + \alpha_\nu T_i(x) = ((A^{\nu-1}B)_i + \alpha_1(A^{\nu-2}B)_i + \cdots + \alpha_{\nu-1}B_i)H(x) + ((A^{\nu-2}B)_i + \cdots + \alpha_{\nu-2}B_i)HF(x)) + \cdots + (B_iHF^{\nu-1}(x))$$

where $\alpha_1, \alpha_2, \dots, \alpha_\nu$ are the coefficients of the characteristic polynomial of the matrix A.

At the same time the mapping T(x) must satisfy (4.3) and we can calculate

$$qF^\nu(x) + \alpha_1qF^{\nu-1}(x) + \cdots + \alpha_\nu q(x) = (CA^{\nu-1}B + \alpha_1CA^{\nu-2}B + \cdots + \alpha_{\nu-1}CB + \alpha_\nu D)H(x)$$

$$+(CA^{\nu-2}B + \cdots + \alpha_{\nu-2}CB + \alpha_{\nu-1}D)HF(x) + \cdots + (CB + \alpha_1D)HF^{\nu-1}(x) + DHF^\nu(x)$$

That is,

$$qF^\nu(x) + \alpha_1qF^{\nu-1}(x) + \cdots + \alpha_\nu q(x) = \beta_0HF^\nu(x) + \beta_1HF^{\nu-1}(x) + \cdots + \beta_{\nu-1}HF(x) + \beta_\nu H(x) \quad (5.1)$$

where

$$\begin{aligned} \beta_0 &= D \\ \beta_1 &= CB + \alpha_1D \\ \beta_2 &= CAB + \alpha_1CB + \alpha_2D \\ &\vdots \\ \beta_{\nu-1} &= CA^{\nu-2}B + \cdots + \alpha_{\nu-2}CB + \alpha_{\nu-1}D \\ \beta_\nu &= CA^{\nu-1}B + \alpha_1CA^{\nu-2}B + \cdots + \alpha_{\nu-1}CB + \alpha_\nu D \end{aligned} \quad (5.2)$$

Which proves that $qF^{\nu-1}(x) + \alpha_1qF^{\nu-1}(x) + \cdots + \alpha_{\nu-1}qF(x) + \alpha_\nu q(x)$ is $\mathbb{R}$-linear combination of $H_j(x), H_jF(x), \dots, H_jF^\nu(x), j = 1, \cdots, p$,

**Sufficiency:** Suppose that there exist constant row vectors $\beta_0$, $\beta_1$, …, $\beta_{\nu-1}$, $\beta_\nu$ that satisfy (5.1). Consider the following choices of (A, B, C, D) matrices:

$$A = \begin{bmatrix} 0 & 0 & \cdots & 0 & -\alpha_\nu \\ 1 & 0 & \cdots & 0 & -\alpha_{\nu-1} \\ \vdots & \vdots & \ddots & \vdots & \vdots \\ 0 & \cdots & 1 & 0 & -\alpha_2 \\ 0 & \cdots & 0 & 1 & -\alpha_1 \end{bmatrix}, \quad B = \begin{bmatrix} \beta_\nu - \alpha_\nu\beta_0 \\ \beta_{\nu-1} - \alpha_{\nu-1}\beta_0 \\ \beta_{\nu-2} - \alpha_{\nu-2}\beta_0 \\ \vdots \\ \beta_1 - \alpha_1\beta_0 \end{bmatrix},$$

$$C = [0\ 0 \cdots 0\ 1]$$
$$D = \beta_0 \quad (5.3)$$

For the above A and C matrices (in observer canonical form), the design conditions (4.2) and (4.3) can be written component-wise as follows:

$$T_1F(x) + \alpha_\nu T_\nu(x) - B_1H(x) = 0 \quad (5.4)$$

$$T_2F(x) - T_1(x) + \alpha_{\nu-1}T_\nu(x) - B_2H(x) = 0 \quad (5.5)$$

$$\vdots$$

$$T_\nu F(x) - T_{\nu-1}(x) + \alpha_1T_\nu(x) - B_\nu H(x) = 0 \quad (5.6)$$

$$T_\nu(x) + DH(x) = q(x) \quad (5.7)$$

We observe that the above equations are easily solvable sequentially for $T_\nu(x), T_{\nu-1}(x), \dots, T_1(x)$, starting from the last equation and going up. In particular, for the chosen B and D matrices, we find from (5.7), (5.6), … , (5.5):

$$T_\nu(x) = -\beta_0H(x) + q(x)$$

$$T_{\nu-1}(x) = -\beta_0HF(x) - \beta_1H(x) + qF(x) + \alpha_1q(x)$$

$$\vdots$$

$$T_1(x) = -\beta_0HF^{\nu-1}(x) - \cdots - \beta_{\nu-2}HF(x) - \beta_{\nu-1}H(x) + qF^{\nu-1}(x) + \alpha_1qF^{\nu-2}(x) + \cdots + \alpha_{\nu-1}q(x)$$

whereas (5.4) gives:

$$\beta_0HF^\nu(x) + \beta_1HF^{\nu-1}(x) + \cdots + \beta_{\nu-1}HF(x) + \beta_\nu H(x) = qF^\nu(x) + \alpha_1qF^{\nu-1}(x) + \cdots + \alpha_\nu q(x)$$

which is exactly (5.1). Thus, we have proved that

$$T(x) = \begin{bmatrix} \left( \begin{matrix} -\beta_0HF^{\nu-1}(x) - \cdots - \beta_{\nu-2}HF(x) - \beta_{\nu-1}H(x) + \cdots \\ +qF^{\nu-1}(x) + \alpha_1qF^{\nu-2}(x) + \cdots + \alpha_{\nu-1}q(x) \end{matrix} \right) \\ \vdots \\ -\beta_0HF(x) - \beta_1H(x) + qF(x) + \alpha_1q(x) \\ -\beta_0H(x) + q(x) \end{bmatrix} \quad (5.8)$$

satisfies the design conditions (4.2) and (4.3) when $\beta_0$, $\beta_1$, …, $\beta_{\nu-1}$, $\beta_\nu$ satisfy (5.1) and the A, B, C, D matrices are chosen according to (5.3).

It is important to observe that the sufficiency part of the proof is constructive: it gives an explicit solution of the design equations (4.2) and (4.3) in terms of the vectors $\beta_0$ , $\beta_1$, …, $\beta_{\nu-1}$, $\beta_\nu$ that satisfy (5.1).

## VI. LOWER ORDER FUNCTIONAL OBSERVERS FOR LINEAR SYSTEMS

The results of the previous section can now be specialized to linear time-invariant systems. The following is a Corollary to Proposition 1.

Proposition 2: For a linear time-invariant system of the form

$$x(k+1) = Fx(k) \qquad (2.6)$$
$$y(k) = Hx(k)$$
$$z(k) = qx(k)$$

there exists a functional observer of the form

$$\hat{\xi}(k+1) = A\hat{\xi}(k) + By(k) \qquad (2.7)$$
$$\hat{z}(k) = C\hat{\xi}(k) + Dy(k)$$

with the eigenvalues of A being the roots of a given polynomial $\lambda^\nu + \alpha_1\lambda^{\nu-1} + \cdots + \alpha_{\nu-1}\lambda + \alpha_\nu$,
if and only if

$$(qF^\nu + \alpha_1 qF^{\nu-1} + \cdots + \alpha_{\nu-1}qF + \alpha_\nu q)$$
$$\in \mathrm{span}\{H_j, H_jF, \dots, H_jF^\nu, j = 1, \cdots, p\} \qquad (6.1)$$

The above Proposition provides a simple and easy-to-check feasibility criterion for a lower-order functional observer with a pre-specified set of eigenvalues governing the error dynamics. Moreover, an immediate consequence of the Proposition 2 is the following:

Corollary: Consider a linear time-invariant system of the form (2.6) with observability index $\nu_o$. Then, there exists a functional observer of the form (2.7) of order $\nu = \nu_o - 1$ and arbitrarily assigned eigenvalues.

The result of the Corollary, derived through a different approach, is exactly the discrete-time version of Luenberger's result for functional observers for continuous linear time-invariant systems [1, 2].

## VII. CASE STUDY

Consider a Continuous Stirred Tank Reactor (CSTR) where N-pyridine is oxidized through Hydrogen Peroxide under non-isothermal conditions[12].

The discretized model equations are as follows

$$C_A(k+1) = C_A(k) + \delta_t\left(\frac{F}{V}\left(C_{A,in} - C_A(k)\right) - R\left(C_A(k), C_B(k), \theta(k)\right)\right)$$

$$C_B(k+1) = C_B(k) + \delta_t\left(\frac{F}{V}\left(C_{B,in} - C_B(k)\right) - R\left(C_A(k), C_B(k), \theta(k)\right)\right)$$

$$\theta(k+1) = \theta(k) + \delta_t\left(\frac{-\Delta H_R}{\rho c_p} R\left(C_A(k), C_B(k), \theta(k)\right)\right) + \delta_t\left(\frac{F}{V}\left(\theta_{in} - \theta(k)\right) - \frac{US_A}{\rho c_p V}\left(\theta(k) - \theta_j(k)\right)\right) \qquad (7.1)$$

$$\theta_j(k+1) = \theta_j(k) + \delta_t\left(\frac{F_j}{V_j}\left(\theta_{j,in} - \theta_j(k)\right) + \frac{US_A}{\rho_j c_{p_j} V_j}\left(\theta(k) - \theta_j(k)\right)\right)$$

$$y_1(k) = \theta(k)$$

$$y_2(k) = \theta_j(k)$$

where the state vector $x = [C_A, C_B\ \theta, \theta_j]$ consists of N-methyl pyridine concentration, hydrogen peroxide concentration, reactor temperature and coolant temperature of the outlet. $\theta$ and $\theta_j$ are the outputs and are assumed to be measurable. $\delta_t$ is the sampling period. In the above equations, $R(C_A, C_B, \theta) = \frac{A_1 e^{-\frac{E_1}{\theta}} A_2 e^{-\frac{E_2}{\theta}} C_A C_B Z}{1 + A_2 e^{-\frac{E_2}{\theta}} C_B} + A_3 e^{-\frac{E_3}{\theta}} C_A C_B$ is the reaction rate and

$C_{A,in}, C_{B,in}, \theta_{in}, \theta_{j,in}$ represent the inlet values of the state values. $F$ and $F_j$ are the inlet feeds and coolant flowrates respectively. $V$ and $V_j$ are the reactor volume and cooling jacket volume respectively. Parameters $A_1, A_2, A_3$ are the pre-exponential factors in the reaction rate. $\Delta H_R$ is the enthalpy of the reaction. $\rho$, $c_p$ and $\rho_j, c_{p_j}$ are the densities and heat capacities of the reactor contents and cooling fluid respectively.

The parameter values are as follows, $C_{A,in} = 2\,\frac{mol}{L}, C_{B,in} = 1.5\,\frac{mol}{L}, \theta_{in} = 373\ K, \theta_{j,in} = 300\ K, \delta_t = 0.5\ s, F = 0.1\,\frac{L}{min}, F_j = 1\,\frac{L}{min}, V = 0.5\ L, V_j = 3 \times 10^{-2} L, A_1 = e^{8.08} L\ mol^{-1} s^{-1}, A_2 = e^{28.12} L\ mol^{-1} s^{-1}, A_3 = e^{25.12} L\ mol^{-1}. \Delta H_R = -160\,\frac{kJ}{mol}, \rho = 1200\,\frac{g}{L}, \rho_j = 1000\,\frac{g}{L}$ $c_{p_j} = 3\,\frac{J}{gK}$. $c_p = 3.4\frac{J}{gK}$ $U = 0.942\ \frac{W}{m^2K}, S_A = 1\ m^2$, $Z = 0.0021\,\frac{mol}{L}$, $E_1 = 3952$ K, $E_2 = 7927$ K, $E_3 = 12989$ K.

Our goal is to design a functional observer that tracks the total concentration of the reactants $z(k) = C_A(k) + C_B(k)$ in the reactor.

The initial condition of the reactor is $C_A(0) = 0, C_B(0) = 0, \theta(0) = 300, \theta_j(0) = 300$ and the model is can be converted to deviation form $C_A' = C_A - C_{A,ref}, C_B' = C_B - C_{B,ref}\ \theta' = \theta - \theta_{ref}$, $\theta_j' = \theta_j - \theta_{j,ref}$ and $y_1' = \theta'$, $y_2' = \theta_j'$ where the subscript ref denotes reference steady state value, with $C_{A,ref} = 0.6684\,\frac{mol}{L}, C_{B,ref} = 0.1684\,\frac{mol}{L}, \theta_{ref} = 410.2332, \theta_{j,ref} = 302.03384$

A scalar functional observer is built ($\nu$ =1) and the necessary and sufficient condition (5.1) is satisfied for the following choice of $\beta_0, \beta_1 \in \mathbb{R}^2$ and $\alpha_1 \in \mathbb{R}$

$$\beta_0 = \left[-\frac{2\rho c_p}{-\Delta H_R}, 1\right]$$

$$\beta_1 = [\frac{2\rho c_p}{-\Delta H_R}\left(1 - \frac{F\delta_t}{V} - \frac{US_A\delta_t}{\rho c_p V}\right) - \frac{US_A\delta_t}{\rho_j c_{pj} V_j},$$

$$\frac{F_j\delta_t}{V_j} + \frac{US_A\delta_t}{\rho_j c_{pj} V_j} + \frac{2US_A\delta_t}{-\Delta H_R V} - 1]$$

$$\alpha_1 = \frac{\delta_t F}{V} - 1$$

Remark 7.1: A sampling period $\delta_t$ that satisfies $\delta_t < 2\left(\frac{V}{F}\right)$ ensures $-1 < \alpha_1 < 1$.

The resulting functional observer is

$$\hat{\xi}(k+1) = -\left(\frac{\delta_t F}{V} - 1\right)\hat{\xi}(k)$$
$$-\delta_t\left[\frac{2US_A}{(-\Delta H_R)V} + \frac{US_A}{\rho_j c_{pj} V_j}\right]y_1'(k)$$
$$+\delta_t\left[\frac{F_j}{V_j} - \frac{F}{V} + \frac{US_A}{\rho_j c_{pj} V_j} + \frac{2US_A}{-\Delta H_R V}\right]y_2'(k) \qquad (7.2)$$

$$\hat{z}(k) = \hat{\xi}(k) - \frac{2\rho c_p}{(-\Delta H_R)}y_1'(k) + y_2'(k)$$

The estimate generated by the functional observer (in non-deviation form) and the estimation error plotted in figure 1.

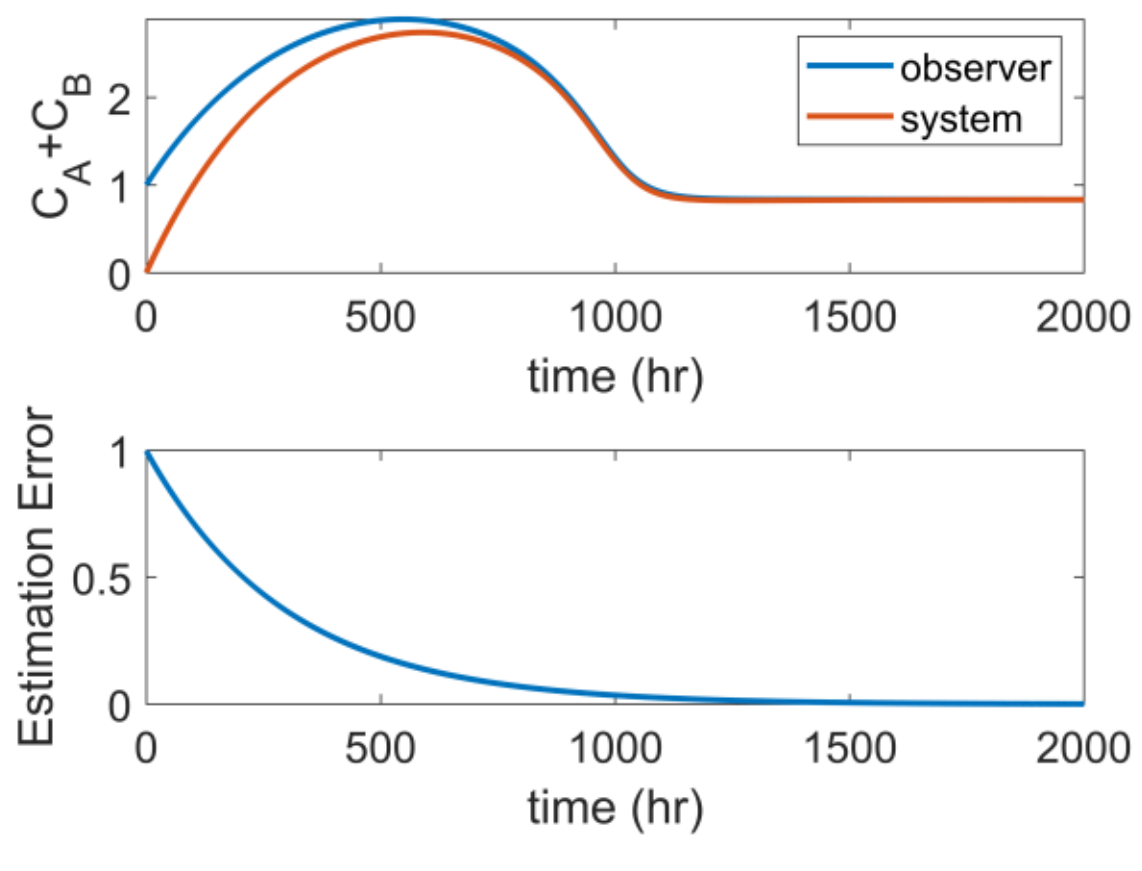

Figure 1: Top-Estimates and true profiles in non-deviation form in the presence of initialization error ($\hat{\boldsymbol{\xi}}(\mathbf{0}) - \mathbf{T}(\mathbf{x}(\mathbf{0}))$=1) where $\mathbf{T}(\mathbf{x})$ is given by (5.8). Bottom-Estimation error ($\hat{\mathbf{z}}(\mathbf{k}) - \mathbf{z}(\mathbf{k})$)

## VIII. Conclusions

A generalization of Luenberger's functional observer to the discrete-time nonlinear systems is presented in this work. The problem of exact linearization of the functional observer dynamics has been studied and conditions for the linearization to be feasible have been derived including a simple formula for the design of the resulting functional observer.